\documentclass[useAMS,usenatbib]{mn2e}
\usepackage{graphicx}

\newcommand{\msun}{M$_\odot$}

\title[V605 Aquilae: a born again star, a nova or both?]{V605 Aquilae: a born again star, a nova or both?}
\author[H.~B. Lau, O. De Marco, X.~W. Liu]{Herbert H.~B.
Lau$^{1,2}$\thanks{herbert.lau@monash.edu.au}, Orsola De Marco$^3$,  X-W Liu$^{2,4}$ \\
$^1$Centre for Stellar and Planetary Astrophysics, School of Mathematics,
Building 28, Monash University, Clayton VIC 3800, Australia\\
$^2$Kavli Institute for Astronomy \& Astrophysics, Peking University, Beijing 100871, P.R.China\\
$^3$Department of Physics \& Astronomy, Macquarie University, Sydney NSW 2109 Australia\\
$^4$Department of Astronomy, Peking University, Beijing 100871, P.R.China}
\begin{document}
\bibliographystyle{mn2e}

\date{Accepted 2010 August 19. Received 2010 June 28}

\pagerange{\pageref{firstpage}--\pageref{lastpage}} \pubyear{0000}

\maketitle

\label{firstpage}

\begin{abstract}
V605 Aquilae is today widely assumed to have been the result of a final helium shell flash occurring on a single post-asymptotic giant branch star. The fact that the outbursting star is in the middle of an old planetary nebula and that the ejecta associated with the outburst is hydrogen deficient supports this diagnosis. However, the material ejected during that outburst is also extremely neon rich, suggesting that it derives from an oxygen-neon-magnesium star, as is the case in the so-called neon novae. We have therefore attempted to construct a scenario that explains all the observations of the nebula and its central star, including the ejecta abundances. We find two scenarios that have the potential to explain the observations, although neither is a perfect match. The first scenario invokes the merger of a main sequence star and a massive oxygen-neon-magnesium white dwarf. The second invokes an oxygen-neon-magnesium classical nova that takes place shortly after a final helium shell flash. The main drawback of the first scenario is the inability to determine whether the ejecta would have the observed composition and whether a merger could result in the observed hydrogen-deficient stellar abundances observed in the star today. The second scenario is based on better understood physics, but, through a population synthesis technique, we determine that its frequency of occurrence should be very low and possibly lower than what is implied by the number of observed systems. While we could not envisage a scenario that naturally explains this object, this is the second final flash star which, upon closer scrutiny, is found to have hydrogen-deficient ejecta with abnormally high neon abundances. These findings are in stark contrast with the predictions of the final helium shell flash and beg for an alternative explanation.
\end{abstract}


\section{Introduction}
\label{sec:introduction}

In 1919 a star was noticed to brighten in the constellation of Aquila \citep{Wolf1920}. A spectrum of the central source revealed it to be a hydrogen-deficient giant \citep{Lundmark1921}. Shortly after the nova-like outburst, V605~Aql brightened over a period of 2 years to a peak of $m_{pg}=10.2$ in 1919. The surface temperature of the star was $\sim$5,000 K, and its spectrum in 1921 was very similar to that of R Coronae Borealis (RCB) stars, hydrogen-deficient, helium-rich supergiants \citep[]{Clayton96,Clayton97}. V605~Aql was later noticed to be in the middle of an old and faint planetary nebula (PN; \citealt{Bidelman1971}), Abell~58 (A~58), with a dynamical age of 20\,000~yr \citep{Pollacco92}. A bright hydrogen-deficient knot was noticed growing at its geometric centre by \citet{Seitter1987}. \citet{Clayton06} estimated today's surface temperature of the central star to be 95,000K. The stellar spectrum and chemical abundances place this central star of PN in the Wolf-Rayet spectral class (also called [WR] by \citet{vanderHucht1981}, to distinguish these stars from their massive counterparts). 

Today, the explanation for the observations of A~58 and its central star is that the central star, after the formation of its surrounding PN, underwent a very late helium shell flash, also called a final flash, which ejected freshly processed stellar material into the centre of the nebula \citep{Iben83, Herwig01}. V605~Aql is considered an older twin of Sakurai's object that underwent a similar outburst in 1995 \citep{Nakano1996}. Other PN consisting of (or containing) hydrogen-deficient ejecta have been found (e.g., A~30 and A~78; \citealt{Jacoby1983}) and they too are considered final flash objects. The extent of the hydrogen deficiency of their stellar atmospheres or ejecta is explained by the timing of the final flash \citep{Herwig01}.

The central star of A~58 has a Wolf-Rayet spectral type, a class that comprises about 15\% of all central stars. Atmospheric abundances of these stars show carbon and helium with a small percentage of oxygen. The atmospheric mass fractions of V605~Aql are C=0.40, He=0.54 and O=0.05 \citep{Clayton06} and they are typical of the inter-shell region and in line with predictions of a post-final flash star \citep{Werner06}. 

Observations of V605~Aql and its surrounding PN therefore seem to agree with final flash models. There is however one observational measurement that is so glaringly in disagreement with final flash models that demands a reconsideration of the final flash model as applied not only to V605~Aql but also the other objects in this class.  \citet{Wesson08} determined accurate abundances of the inner hydrogen-deficient nebulosity in A~58 and found C/O=0.06 and a neon mass fraction of 0.34. They argued that these abundance patterns  have more in common with oxygen-neon-magnesium (ONeMg) nova ejecta than the predictions of the final flash.  The very low C/O ratio observed in the hydrogen-deficient ejecta inside the A~58 PN is hard to reproduce in stellar models,  for example, \citet{Karakas09} showed that the C/O ratio in a planetary nebula can be lower than unity, but no lower than 0.38 by number ($\sim 0.29$ by mass). Even more importantly, the extremely high neon abundances cannot be reproduced in any final flash model. The only way to obtain such high neon abundance is to dredge neon up from a neon core.


In addition two of the hydrogen deficient knots in the centre of the old PN A~30, also thought to be a final flash star, were measured by \citet*{Wesson2003} to have C/O=0.19 and 0.21, respectively and neon mass fractions of 0.08 and 0.20, respectively.
It therefore appears that two of the handful of known final flash stars do not fit the final flash scenario.

For these reasons we investigate possible scenarios that invoke an ONeMg white dwarf (WD) as the cause of the hydrogen-deficient ejecta in A~58, while at the same time explaining the other observed characteristics.


In \S~\ref{sec:scenarios} we construct plausible scenarios and outline their drawbacks. In \S~\ref{sec:BSEcode} we explain the population synthesis code that was used to determine the likelihood of one of the proposed scenarios, and describe the population synthesis results. We conclude in \S~\ref{sec:conclusion}.

\begin{table*}\def~{\hphantom{0}}
\begin{center}
\caption{Abundance comparison}
  \label{tab:abundances}
\begin{tabular}{lccccccl}
\hline
Name/Type & \multicolumn{6}{c}{Mass Fractions (\%)} & Reference\\
                  & H & He & C & N & O & Ne & \\
                  \hline
A~58 knot   & 2 & 25 & 2 & 4 & 32 & 35 & \citealt{Wesson08}\\
A30-J1 knot & 1 & 52 & 7 & 5 & 27 & 8 &  \citealt{Wesson2003}\\
A30-J2 knot & 1 & 49 & 6 & 4 & 21 & 20 & \citealt{Wesson2003} \\
\hline
Nova prediction &27-33 & 16-22 & 0.6-4$^c$ & 2-9$^d$ &7-12$^e$ &17-25& \citealt{Starrfield1998}\\
V693~CrA Ne nova &41 & 21 & 0.4 & 7 & 7  & 24 & \citealt*{Vanlandingham1997}\\ 
V4160~Sgr Ne nova & 47 & 34 & 0.6 & 6 & 6 & 7 & \citealt{Schwarz2007}\\ 
V1370~Aql Ne nova & 5 & 9 & 3 & 14 & 5 & 52 & \citealt{Snijders1987}\\
QU~Vul Ne nova& 33 & 26 & 1 & 7 & 17 & 9 & \citealt*{Andrea1994}\\
\hline
final flash prediction & - & $\sim$30 & $\sim$45$^b$ & $<$3$^a$ & $\sim$20$^b$ & $\sim$2 & \citealt{Werner06}\\
V605~Aql star & - & 54 & 40 & - & 5 & - & \citealt{Clayton06}\\
PG1159-035 star & $\leq$2 & 33 & 48 & 0.1 & 17 & 2 & \citealt{Jahn2007}\\
\hline
\multicolumn{8}{l}{$^a$We assumed this number from the prediction of 1 to a few percent of \citet{Werner06}.}\\
\multicolumn{8}{l}{$^b$We assumed these average numbers from different predictions cited in \citet{Werner06}.}\\
\multicolumn{8}{l}{$^c$This is by summing $\rm ^{12}C$ and $\rm ^{13}C$.}\\
\multicolumn{8}{l}{$^e$This is by summing $\rm ^{14}N$ and $\rm ^{15}N$.}\\
\multicolumn{8}{l}{$^e$This is by summing $\rm ^{16}O$ and $\rm ^{17}O$.}\\

\end{tabular}
 \end{center}
\end{table*}

\begin{table*}\def~{\hphantom{0}}
  \begin{center}
  \caption{Initial conditions and comparison with observations for two binary scenarios.}
  \label{tab:scenarios}
  \begin{tabular}{lll}
  \hline
     &Merger   scenario   &Nova scenario \\
 \hline    
$M_1$  & massive AGB & ONeMg WD\\
$M_2$  & main sequence star &  AGB\\
Separation& $<$few AU & a few hundred AU\\
 \hline  
Observations  & & \\                      
Old PN (borderline Type~I)  &    Accounted             & Type I or non-Type I, depending on secondary mass\\
Old PN elliptical       & Accounted? & Accounted\\
Final flash: 1917                     & is the merger event & Final flash from secondary\\
H-deficient giant 1921       & produced by the merger&from final flash\\
O- and Ne-rich ejecta & from the ONeMg primary & from nova right after final flash\\
$[WC]$ 1987-today      & evolution of post-merger object& regular evolution of the post final flash star\\
\hline
   \end{tabular}
 \end{center}
\end{table*}

\section{Possible scenarios}
\label{sec:scenarios}


The premise under which we operate is that  the presence of neon- and oxygen-rich ejecta came about because of an explosion that involved a massive, ONeMg WD. ONeMg WDs derive from super-asymptotic giant branch (SAGB) stars, stars hot enough to ignite carbon in the early AGB phase resulting in an ONeMg core. An extensive set of SAGB models can be found, e.g., in \citet{Doherty10}. Additional requirements derived from observations are that the outburst took place shortly after departure from the SAGB and the ejection of a regular PN. In addition, after the outburst the star became a hydrogen-deficient supergiant in only a few years (1917-1921), after which it heated up and developed a [WC] spectral type (taking no longer than $\sim$65 years, \citealt{Clayton06}). Moreover, we need to explain the event that led to the 1917 outburst itself. An obvious scenario is that of a nova that went off in 1917-1919 and that was misinterpreted as a final flash. The immediate problem with this scenario is that a nova is unlikely to produce a hydrogen-deficient star, although one neon nova with abnormally low abundances of hydrogen in its ejecta is actually known \citep[V1370 Aql;][]{Snijders1987; Table 1}. According to \citet{Austin96}, the observed high neon abundances of this nova can only be explained by the ejection of core material from a ONeMg white dwarf and the estimated mass of the white dwarf is $1.3$~M$_\odot$. A second problem with a nova scenario to explain the 1917-1919 outburst of V605~Aql is that the light curve behaviour of this object is not that of a neon nova: V605~Aql remained a giant after the outburst for at least $\sim$50 years. Its dust production as well as its current spectral type also do not match the behaviour of post-nova stars. These are the reason why we have searched for alternative scenarios.
The best two scenarios are described below and summarised in Table~\ref{tab:scenarios}, along with the observations they attempt to explain.

\subsection{The common envelope merger scenario}
\label{ssec:ce1}

As we have described, after the outburst V605~Aql became a hydrogen deficient giant. Its spectrum and variability behaviour was that of an RCB star \citep{Clayton97} and very similar to that of Sakurai's object \citep[e.g.,][]{Tyne02}. We do not know what the exact abundance of V605~Aql at that time was, as that spectrum could not be reliably modelled, but we do know that RCB stars are primarily made of helium. Since  we know that RCB stars are likely to result from a merger \citep{Clayton07}, we have constructed a merger scenario for  V605~Aql even if abundances of the post-merger RCB stars are dominated by helium rather than {\it carbon} and helium as is the case for V605~Aql today \citep{Clayton07}. This said, since mergers remain quite complex phenomena, we will relax this constraint and assume that there is a way to make a [WC] star with a merger \citep[see also][]{DeMarco2002}. 

In this scenario a massive AGB star ($M \la 6-8$~M$_\odot$) suffers a common envelope \citep{Paczynski1976} with a lower mass main sequence star, when the primary evolves to the SAGB. This common envelope results in a merger which strips the primary of hydrogen (by ejection and ingestion) revealing the intershell region. If the primary is massive enough to have an ONeMg core, then the ejecta would be rich in neon and have a C/O ratio lower than unity, as observed. 

One way in which a single common envelope event would result in a regular, hydrogen-rich PN {\it and} in hydrogen-deficient ejecta inside the main PN is if the common envelope interaction takes place over a much longer time than the few years envisaged by hydrodynamic simulations \citep{Sandquist1998,DeMarco03}.  \citet{Baer2010} envisaged such a slow common envelope where the companion lingers on the outskirts of the primary inducing mass-ejection. Only later does it plummets towards the core in a faster phase of in-spiral. In such a scenario the first PN would derive from early ejection of regular envelope material (the pre-common-envelope mass loss), while the hydrogen-deficient ejecta would derive from the final in-spiral and merger. 

A main criticism of this scenario is that a PN becomes visible because post-AGB stellar wind ploughs up the material ejected during the AGB phase. In this scenario by the time the star is in the post-AGB phase and able to plough up material, both the hydrogen-rich and hydrogen-deficient ejecta have been expelled. This is contrary to what is implied by the two nebulae, where the first, larger one must have been plowed up by the post-AGB wind {\it before} the ejection of the second, hydrogen-deficient one. To fix this inconsistency we envisage a variation of this scenario where the first common envelope resulted in a regular, hydrogen-rich PN approximately 20\,000 years ago. The envelope of the primary was slowly stripped off through this slow common-envelope process. The post-common envelope object was a short-period binary where the primary is an ONeMg WD and the secondary a low-mass main-sequence star. Eventually the WD suffers a final flash. The resulting expansion forms a new common envelope with the nearby secondary. This second common envelope results in a merger. The ejected second common envelope is hydrogen deficient and neon rich.

A SAGB star might be expected to eject a Type I PN, or a PN heavily enriched in nitrogen by the hot-bottom burning process (N/O=0.8, by number; \citealt{Kingsburgh1994}). A~58 has N/O=0.78 \citep{Guerrero96}, very close to the Type I limit. Finally, the shape of the old PN is not what one might expect of a post-common envelope PN, which tend to be bipolar \citep[although they are not exclusively bipolar;][]{Miszalski2009}. 
A possible drawback of this scenario is that the mass of today's [WR] central star would be expected to be relatively large \citep[$\sim$0.95~M$_\odot$][]{Weidemann2000}, while the mass of V605~Aql today appears to be lower ($\sim$0.61~M$_\odot$; \citealt{Lechner2004}).

In this scenario the witnessed 1917-1919 outburst would be the common envelope event (the timescale for the existence of the giant would not be dissimilar from those of a common envelope event -- one to a few decades \citep{Sandquist1998,DeMarco03}, if we adopt the single-common envelope scenario. In the double-common envelopes scenario the outburst would be the final flash. We note that in this scenario the neon ejecta do not derive from a nova type outburst, but rather from a merger with a massive, ONeMg WD.

\subsection{The nova scenario}
\label{ssec:nova}

In this scenario we have an ONeMg WD primary and an AGB secondary (which can have any mass). The separation is intermediate. Mass transfer is taking place from the AGB wind (not from Roche lobe overflow) onto the WD primary. The AGB secondary makes a PN by regular evolution and becomes a WD. During this time, mass transfer is reduced greatly since the secondary radius is smaller and its mass-loss is reduced.  Eventually, the WD secondary experiences a final flash that produces carbon-rich ejecta (these ejecta would have a C/O ratio larger than unity and a very low neon abundance). During the final flash, the new hydrogen-deficient giant expands and starts transferring mass to the ONeMg WD primary again. This renewed accretion pushes the WD  over the limit for a nova detonation. Depending on the accretion rate, the nova can occur relatively quickly. The nova ejecta is oxygen and neon rich and becomes mixed with the recently produced final flash ejecta. The primary WD eventually fades as a typical post-nova WD, while the secondary WD follows the canonical post-final flash evolution, developing a [WC] spectral type. 

The observed hydrogen-deficient ejecta are, in this scenario, a mix of the final flash and nova ejecta. As we can see from Table~\ref{tab:abundances}, it is not impossible to conceive of a scenario that mixing of these two types of ejecta would result in the observed abundances, as long as the ejected masses are approximately similar. This is likely to be the case: the hydrogen-deficient ejecta mass of A~58 was measured to be 5.25$\times 10^{-5}$~\msun \citep{Wesson08}, while nova ejecta masses could be in the range 1-10$\times 10^{-6}$ \citep{Starrfield1998}. If today's [WR] central star is the secondary, final flashes can explain the surface temperature change from 5000K in 1919 to 95,000~K now. This scenario would also explain the difference in the ejecta and the stellar abundances: the massive WD primary is responsible for the observed abundances of the hydrogen-deficient knot, while the secondary is responsible for the observed central star abundances. 



The nova explosion would, in this scenario,  follow shortly after the final flash. We therefore wonder whether the nova outburst should have been detected, as was the case for the outburst due to the final flash. As it turns out, monitoring of this star since 1917 has been relatively sparse and ONeMg novae tend to be relatively dim and return to light minimum in relatively short time scales. The lowest apparent peak $V$ magnitude of novae observed in the LMC is $\sim$12.5 \citep{Shida04}. Scaling this magnitude to the distance of V605~Aql ($\sim$3.5 kpc, \citealt{Clayton97}), the apparent (dereddened) $V$ magnitude of our nova might have been  $\sim$6.8. If we take into account an interstellar reddening $\rm{A_v} = 1.7$ \citep[for a full discussion of the reddening see][]{Clayton97} the apparent magnitude could have been as low as 8.5 mag. The lightcurve  reported by \citet{Harrison96}, between 1917 and 1924 shows that V605~Aql reached a peak of $ m_{pg}\sim 10$ but later remained between 12 and 14 magnitudes. A nova with an apparent magnitude of 9, might have been easily detected, were it not for the fact that the light curve of V605~Aql was sparsely sampled and a nova could have gone off and returned to minimum light between observations. If the accretion rate at the time of the final flash was high enough and the WD was massive, the neon nova decline time could be as short as 12 days and its luminosity would be a meagre $\sim 4.8-6\times10^{4}\rm L_\odot$ \citep{Prialnik95}. It is therefore not excluded that a nova did indeed take place but remained undetected. We also note that if we require a dim nova with a fast return to light minimum, other nova explosions must have occurred in the past of this system. Hence, it is possible that other knots with similar abundances could be found within the PN, though they will be further away from the central star. It would also follow that other outbursts may yet be detected.

\section{Population synthesis test}
\label{sec:BSEcode}

Of the two scenarios, the nova one described in \S~\ref{ssec:nova} is the only one where each phase is reasonably well understood and we can therefore apply a stellar population synthesis model to determine its frequency of occurrence. To determine the frequency of such systems, we use a rapid binary-evolution algorithm \citep*[BSE; ][]{Hurley02}. Using this code, we can generate a binary population and determine which systems can lead to the formation of A~58-like systems.
The code uses the detailed single-star evolution formulae of \citet*{Hurley00} to calculate the stellar luminosity, radius,
core mass, core radius and spin frequency for each of the component stars as they evolve. 
A prescription for common envelope evolution is also included. The $\alpha_{\rm CE}$ parameter, the efficiency of the orbital energy transfer to the envelope during the common envelope evolution, is set to be equal  to unity. This parameter is very uncertain.

Details of the binary-evolution algorithm are described by \citet{Hurley02}. Here, we want to highlight the treatment of wind accretion. This is a key process for the scenario described in \S~\ref{ssec:nova} because for a moderately wide binary that avoids common envelope phase and Roche lobe overflow, mass is transferred through wind accretion instead. When a star loses mass in a stellar wind, its companion can accrete some of the material as it orbits through it. Moreover, the mass loss may be tidally enhanced by the presence of the companion if the secondary is moderately close. A descriptive formula given by \citet{Tout88} is used in the code to calculate the enhanced mass-loss rate. Typically, the mass accretion rate onto the secondary will be significantly tidally enhanced when the radius of the primary reaches 10\% of the Roche Lobe radius. The accretion rate is typically of the order of  $10^{-6}-10^{-9}\,\rm M_\odot$~yr$^{-1}$ in the systems that we are investigating.

In our scenario, the primary is already an ONeMg WD and mass is deposited onto it by wind accretion from an AGB secondary. However, the accretion stops when the secondary evolves into a WD and its mass loss decreases. When the final flash causes the secondary to expand again as it evolves back to the post-AGB track, wind accretion restarts. Renewed  mass transfer can take place for only a brief time since the star will soon shrink again, so the amount of mass accreted during this post-final flash phase cannot be large. 

The code is unable to predict exactly when a nova occurs, but it can trace the amount of mass deposited onto the WD. A few test models are used to simulate the mass accretion after the final flash, with a subroutine forcing the secondary WD to be reborn from the WD cooling track back onto the AGB. Based on the test models and nova evolution models by \citet{Prialnik95}, systems with mass accretion rates reaching $10^{-7}\rm M_\odot$~yr$^{-1}$ at the end of AGB evolution will result in a nova shortly after the final flash. The recurrent periods of such novae are $0.771-19.6$~years. The actual required accretion rates could be slightly lower because mass may be deposited on the WD primary before the final flash, but it will not significantly alter our estimated frequency. Our binary system needs to be close enough so that the accretion rate onto the primary WD is high enough to trigger a nova explosion.
 
\begin{figure}
\includegraphics [width=\columnwidth,angle=0]{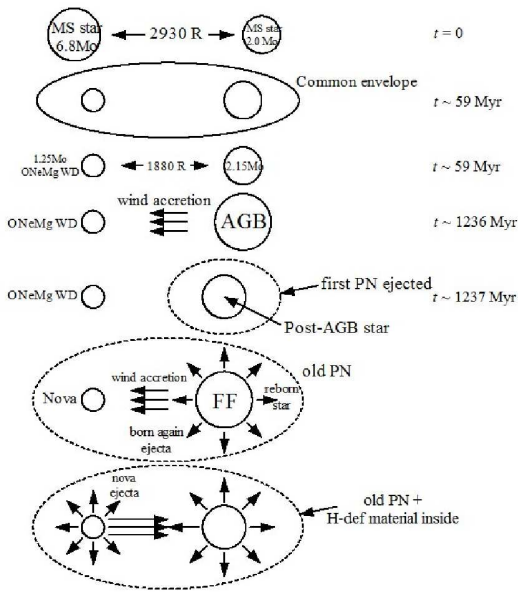}
\caption{A cartoon of an evolutionary path that could lead to A58-like systems. The nova explosion occurs at the end of the evolution after the secondary undergoes a final flash.} 
\label{fig:cartoon}
\end{figure}

As an example of a system that may satisfy our observational constraints, we present here the case of a primary star with main sequence mass of $6.8\rm M_\odot$ and a companion with a main sequence mass of $2\rm M_\odot$; their initial separation is $2930\rm R_\odot$ (see cartoon in Fig.~\ref{fig:cartoon}). The primary star evolves into a SAGB star at the age of  $58.7$ million years and the two stars enter a common envelope phase at an age of $59.5$ million years. After the common envelope phase, the primary has lost all of its envelope and has became an ONeMg WD of mass $1.2$~M$_\odot$ and the secondary is now accreting some of the envelope and has therefore increased in mass to $2.1$~M$_\odot$. The system's separation has been reduced to $1880$~R$_\odot$. Eventually, the secondary evolves to the early AGB phase at a system age of $1.2$ billion years. 
Mass is at this point accreted onto the ONeMg WD at a slow rate of  $10^{-11}\rm M_\odot$~yr$^{-1}$. Eventually the accretion rate increases as the AGB secondary grows in size, reaching a rate of  $10^{-7}\rm M_\odot$~yr$^{-1}$. Since the exact accretion rate during this phase is fundamental to time the nova outburst right after the final flash, the initial separation is crucial.

The lower limit for the separation is set by the requirement that a second common envelope between the AGB secondary and the now WD primary be avoided. This is because the accretion rate before or during a common envelope interaction is extremely uncertain. During the common envelope the accretion rate onto the secondary may be very low, or nonexistent due to the supersonic nature of the companion's orbital speed and the very short timescale of the in-spiral \citep{Sandquist1998,DeMarco03}.  It is possible that accretion may take place at the correct rate before the common envelope phase \citep[e.g.,][]{Baer2010}, but the duration of the phase may be too short. However there is no way at present to quantify this accretion. We therefore consider only systems that avoid a second common envelope and we discuss additional channels with two common envelopes in \S~\ref{sec:conclusion}. 

Under these conditions, the lower separation limit decreases with decreasing secondary mass, because the AGB radius is smaller and the system can avoid the second common envelope phase even with smaller initial separations.  The lower initial mass limit of the secondary is $\sim 0.8\rm M_\odot$, as systems less massive than that do not evolve off the main sequence\footnote{We have not taken into account the fact that the secondary's main sequence mass is increased by accretion during the SAGB phase of the primary and that this will accelerate the secondary evolution. Uncertainties over the exact amount of accretion make it difficult to assess exactly the actual lower limit of the secondary mass. The lower this limit, the larger the number of systems that can evolve via our scenario.}. For a $0.8$~M$_\odot$ secondary and a $6.6$~M$_\odot$ primary, the lowest possible initial separation is $\sim 2000$~R$_\odot$. It is worth pointing out at this stage, that the nitrogen-rich nature of the old PN (ejected, in our scenario, by the secondary), favours the secondary to have a higher initial mass. 

The distribution of initial periods of binary stars is not well known.  It is a common practice to assume the separation to be uniform in logarithmic space \citep{Eggleton89}, based on the observed frequency of doubly bright visual binaries. With this assumption, about 15\% of all systems lie within the range of suitable initial separations. The period distribution of \citet{Duquennoy91}, based on a spectroscopic survey of G-dwarfs, has instead a Gaussian shape, with a median period of 180 yr. If we use this distribution, $\sim$10\% of binary systems lie within the suitable separation range. 

The mass range for stars that evolve to the SAGB is not well known as it depends on different assumptions for convective overshooting on the main sequence \citep{Poelarends08}.  The models used by \citet{Hurley02} do include overshooting and hence give a main-sequence mass range of $6.4-8.1$~M$_\odot$.  

As in \citet{Hurley02}, we assume one binary is born with mass greater than $0.8$~M$_\odot$ in the Galaxy per year. This is in agreement with the WD birth rate in the Galaxy determined by \citet{Phillips89}. Then, based on the initial mass function of \citet{Kroupa93}, only 1 binary system with a primary that will evolve to the SAGB is formed in the Galaxy per century. The distribution of the mass ratio of primary to secondary is uncertain. If we assume a flat distribution of mass ratio, only one out of every 1000 systems will evolve via our scenario each year for the entire Galaxy, or one system per millennium will have the characteristics of A~58 and its central star.

\citet*{Liebert05} have estimated an overall formation rate of WDs in the local Galactic disk to be $1\times 10^{-12}\rm{pc}^{-3}\rm{yr}^{-1}$, which is less than half of the value used in  \citet{Phillips89}. If we adopt this WD formation rate, then 
the frequency of our scenario will halve.

We should further reduce our estimate because not all AGB stars undergo very late thermal pulses. It is estimated that only about 20-25\% of all post-AGB stars become hydrogen-deficient via a final flash \citep{Blocker01}, which will reduce our estimate to one system every 5000-10\,000 years.

\section{Conclusion}
\label{sec:conclusion}

Many questions remain about the final flash scenario and the objects that derive from it. In this paper we have addressed the fundamental issue of the abundance of the ejecta that have supposedly derived from the final flash outburst. These ejecta should be carbon rich (C/O$>$1) and have a very low neon mass fraction ($\sim$2\% by mass; \citealt{Werner06}). They are instead oxygen rich and have extremely large neon mass fractions. These ejecta abundances are more in line with those derived from ONeMg nova ejecta.

We have tried to determine whether a nova scenario can explain these and other observations of A~58 and its central star V605~Aql, but found it difficult to determine a scenario that easily accounts for all the properties of this object. We find two scenarios that {\it could} lead to the observed properties of this system. The first invokes a merger between an ONeMg WD and a main sequence star. In this scenario we have made a broad assumption that such mergers can create hydrogen-deficiency in the resulting star {\it and} eject neon-rich gas from the core of the ONeMg WD. Because of these and other unconstrained assumption, we cannot test the frequency of this scenario.

In the second scenario, the ejecta are partly formed though a nova explosion that closely follows a classical final flash. This is possible but we show, via a population synthesis analysis, this scenario to be quite unlikely (one such is system born every 5000-10\,000 years). If we assume that Sakurai's object is a modern twin of V605~Aql, i.e., that it evolved though the same path and will continue to do so, then we must assume that objects like this are more frequent than what was derived for our nova scenario. There may be other pathways to the formation of A~58 systems (such as a double-common envelope scenario hinted at in \S~\ref{sec:BSEcode}). There is no real advantage at this point, trying to add up the odds of stars evolving through a variety of binary scenarios each of which can only be very approximately modelled. We only remark here on the fact that the actual number of binary channels that can lead to the observed characteristics and timescales of V605~Aql, is not very large.

\section{Discussion}
\label{sec:discussion}

Beside V605~Aql, there are six other systems with hydrogen-deficient ejecta, which have been explained with some variation of the final flash scenario. An eighth system, CK~Vul, may or may not be related to final flash scenario \citep{Hajduk07}. Sakurai's object, appears to be a modern twin of V605~Aql (but the abundances of carbon and neon in its hydrogen deficient nebula have not been determined); its hydrogen deficient material is distributed in a disk \citep{Chesneau09}, as is the case for V605~Aql \citep{Hinkle08}. FG~Sge \citep{Gonzalez98}, was observed to undergo an outburst in the late 1800s. Two other systems, A~30 and A~78, have hydrogen-deficient ejecta inside old round or elliptical PNe \citep{Harrington96}. A~30 is the only other system where the C/O ratio was determined to be sub-unity and the neon mass fraction was very large \citep{Wesson2003}. A~30 and A~78 are broadly regarded to be almost identical systems, both having central stars with the abundances predicted for a final flash \citep[e.g.,][]{Herald05}. Two additional hydrogen-deficient PN are known: IRAS15154-5258 \citep{Zijlstra02} and IRAS18333-2375. IRAS18333-2357 is the only hydrogen-deficient PN which does not reside inside a hydrogen-rich one. \citet{Gillett89} also found this hydrogen-deficient PN to be extremely neon rich. This object is associated with the globular cluster M~22 and is one of the 4 PN known in the globular cluster system of the Galaxy \citep{Jacoby94,Jacoby97}. IRAS18333-2357's hydrogen-deficient PN is very irregularly shaped. Its central star is a reasonably hydrogen-rich O star \citep{Gillett89,Rauch98}, implying the presence of a second undetected star that ejected the hydrogen-deficient material and that is therefore expected to be hydrogen-deficient.

Finally, there is another system we should keep in mind when considering the plausibility of nova scenarios in the context of PN. \citet{Wesson08b} discovered a PN surrounding Nova Vul 2007. \citet{Rodriguez10} measured the period of the central binary to be 0.069~days, the shortest period binary known in a PN. They suggested V458 Vulpeculae  to be a post double common-envelope system, composing a relatively massive white dwarf accreting matter from a post-asymptotic giant branch star which produced the PN observed. The fast nova, with its rapid, 21-day, 3-magnitude brightness decrease from light maximum, suggested the white dwarf mass to be at least $1~M_\odot$.  Despite the clear indication that this system is indeed what it appears, i.e., a nova in a PN, it is difficult to determine a likely scenario for the accretion of mass onto the massive WD at the hand of the secondary: the AGB secondary may have transferred mass onto the WD primary, before the system entered a common envelope. The mass accretion onto the primary during the common envelope phase is likely to have been low, and after the common envelope, when the secondary mass donor detached from its Roche lobe and was shrinking to post-AGB size, one may suppose that accretion stopped altogether. However accretion must have been happening to detonate the nova. It is possible that finding a scenario for this object may illuminate the past of the V605~Aql star.

Despite the successes of the final flash scenario in explaining the varied yet relatively homogeneous characteristics of the seven objects with hydrogen-deficient nebular material, the C/O ratio and neon abundances of the hydrogen-deficient ejecta of two of these seven objects are in glaring disagreement with the final flash scenario and the neon abundances of a third object may also present similar problems. In the light of this disagreement, other characteristics of these systems that are also in disagreement with the final flash scenario, though may be more easily reconciled, acquire renewed interest. We have tried to find a simple binary scenario which, while explaining the abundances also explains the other observations. This has proven difficult, and the scenarios summarised in \S~\ref{sec:conclusion}, while reasonable, seem convoluted. The plot has certainly thickened over the nature of final flash stars.

\bibliography{Abell}

\label{lastpage}

\end{document}